# Thermal Holography with Nonlocal Metasurfaces

Mingze He[1,6], Lin Jing[1,6], Michele Guizzardi[1,2], Shixiong Yin[1], Martina Riva[1,2], Yajun Gao[1], Shuwei Guo[1], Adam C. Overvig[1,3], Sander A. Mann[1,4], Andrea Alù[1,5*]

1. Photonics Initiative, Advanced Science Research Center, City University of New York, New York, NY, 10031, USA
2. Politecnico di Milano, Department of Physics, Milan, Italy.
3. Department of Physics, Stevens Institute of Technology, Hoboken, NJ, 07030, USA
4. Van der Waals-Zeeman Institute, Institute of Physics, University of Amsterdam, Amsterdam, 1098 XH, The Netherlands
5. Physics Program, Graduate Center, City University of New York, New York, NY, 10016, USA
6. These authors contributed equally.

* Email: aalu@gc.cuny.edu

## Abstract

Thermal radiation and luminescence are among the most ubiquitous sources of light, yet their properties are difficult to engineer due to their incoherent nature. Recent advances in thermal emission control have endowed heat-generated light with spatial and temporal coherence, as well as control over its polarization, by judiciously patterning periodic arrays of nanostructures. However, custom wavefront engineering of incoherent light has remained an open challenge, as it requires not only sufficient spatial and temporal coherence but also arbitrary local phase control, which experimentally translates into sustaining long-range lattice resonances in aperiodic systems while maintaining their robustness against local phase perturbations. In this work, we address this longstanding limitation and demonstrate programmable vectorial control over the thermal emission wavefront from single-layer aperiodic nonlocal metasurfaces, realized by manipulating local geometric phases across a collective delocalized resonance. We exemplify the wavefront-shaping capabilities of our approach by demonstrating thermal beam steering, spin-momentum-locked thermal focusing and the generation of a thermal vectorial hologram, providing custom spatial and spectral control over thermal light. Our platform addresses challenges in thermal emission engineering and provides a route for light generation and control without the need for external coherent sources or bulky setups. Future work may extend this design principle to other incoherent emission systems, such as photo- and electroluminescence.

## Introduction

Ubiquitous across nature and technology, thermal emission is intrinsically incoherent, broadband, and unpolarized[1]. Despite these limitations, it remains one of the most cost-effective options for both research laboratories and everyday use[2], especially in the mid-infrared spectrum. However, the incoherent nature of thermal emission poses difficulties for applications where well-defined polarization[3], wavelength[4] and wavefronts[5] are necessary. Similar challenges are shared with other incoherent light sources, such as photoluminescence and electroluminescence. A stack of optical components, e.g., bandpass filters, pinholes, polarizers, waveplates and refractive lenses, can be used to impart coherence and produce narrow-band wavefronts with desired polarization states, but this approach compromises efficiency while increasing the footprint and system complexity. In an effort to simplify and empower thermal light sources[2], temporally and spatially coherent emission has been investigated over the past two decades. A prominent strategy leverages phonon polaritons, which naturally bridge thermal oscillations and mid-infrared light waves[6–15]. By using polaritons, it is possible to enhance spatial coherence[7], and their relatively long lifetime endows temporal coherence[16]. However, these phenomena are limited to specific material resonance frequencies[16], significantly limiting their general applicability and the overall control over thermal light.

Recently, large temporal and/or spatial coherence of thermal light has been demonstrated using dielectric and plasmonic materials[17–25] through structural resonances, offering wider control for applications in the mid-infrared range. In particular, quasi-bound states in the continuum (q-BICs)[26] have been employed to control thermal emission[19–23,27,28], providing large and controllable $Q$ factors supported by collective (nonlocal) resonances in periodic arrays of carefully engineered nanostructures. While q-BICs provide a powerful approach to engineering the temporal coherence of thermal emitters, periodic implementations do not by themselves provide spatially varying wavefront control. At the other extreme, local metasurfaces can spatially structure the wavefront with subwavelength precision, but they rely on coherent illumination to manipulate the phase and amplitude of the scattered fields[29–31]. To control the emission wavefront, a *thermal metasurface* must simultaneously support nonlocal resonances that establish spatial and temporal coherence and local geometric phase control[32,33]. While coherence depends on collective resonances supported by periodicity, local phase control requires breaking this periodicity—a central challenge for programmable control over thermal emission. Diffractive nonlocal metasurfaces[34–41], formed by carefully applying tailored perturbations to an otherwise periodic lattice, provide a route to address this challenge: such structures simultaneously preserve highly delocalized resonant states while tailoring their spatial profiles with subwavelength precision. Theoretical frameworks[32,33] have established the potential of aperiodic architectures for general thermal wavefront shaping. Experimentally, however, prior demonstrations have been limited to polarization-selective unidirectional emission[23] and one-dimensional focusing or quasi-two-dimensional holography synthesized by orthogonally multiplexing one-dimensional grating channels[42], leaving genuine two-dimensional thermal wavefront control unaddressed.

Here, we demonstrate highly coherent thermal emission with custom spatial wavefronts using a single-layer, readily manufacturable leaky-wave metasurface platform based on a q-BIC. The large $Q$ factor of the q-BIC provides temporal coherence, while spatial coherence is maintained through modal nonlocality. We then leverage deliberate local perturbations of BICs, designed to preserve both the high-$Q$ resonance and modal nonlocality while tailoring the local geometric phase within an aperiodically patterned structure. The leaky-wave metasurface[36,38] makes the lattice resonance robust against aperiodicity without requiring local birefringence. Based on these features, we demonstrate programmable control over the thermal emission wavefront, with three functions: beam steering of thermal emission with a linear phase gradient, cylindrical and circular thermal focusing with spin-momentum locking, and a thermal vectorial hologram projecting a desired image at a distance (**Fig. 1a**). Our work, using aperiodically patterned leaky-wave metasurfaces, demonstrates two-dimensional wavefront shaping of thermal emission, providing opportunities for emission engineering and low-cost distributed custom thermal sources.

**Quasi-BIC Engineering**

Thermal wavefront control requires a nonlocal response that establishes coherence together with local geometric-phase control that shapes the emitted field. Hence, we need a periodic lattice structure that supports nonlocal, high $Q$-factor collective resonances (temporal coherence, born of temporal dispersion) with small angular emission (spatial coherence, born of spatial dispersion), while simultaneously enabling manipulation of the local geometric phase without disrupting these nonlocal properties. To satisfy these stringent requirements, we employ a zone-folded q-BIC[43], in which a local geometric phase is introduced through deliberate local perturbations of the BIC mode.

Regarding temporal coherence, in the small perturbation regime, the quality factor of the q-BIC is controlled through the perturbation strength[19,43,44], which can also determine the amplitude, phase and polarization of the leaked fields. This approach does not require the use of polaritonic resonances, yet it can feature $Q$ factors above 100, with thermal emission close to blackbody levels (**SI Section 2**).

Wavefront control also requires spatial coherence, which depends on both the modal lifetime and dispersion. Spectral (temporal) coherence is set by the resonance linewidth, or equivalently the $Q$ factor and modal lifetime, whereas spatial coherence is characterized by the angular width and coherence length. For a mode with parabolic dispersion, the spatial coherence length $L_c$ scales as $\sqrt{\tau b}$ at the band edge, where $\tau$ is the modal lifetime and $b$ is the band curvature[32]. Nonradiative loss ultimately limits the lifetime[19,43,44], while the band structure limits the curvature[20], and together, these factors severely restrict the achievable spatial coherence.

The spatial coherence of a nonlocal mode can be dramatically increased through dispersion engineering. For a mode with linear dispersion, for example, the spatial coherence length is $\tau v_g$, where $v_g$ is the group velocity (**SI Section 3**). This length can readily exceed the spatial coherence length of a band with parabolic dispersion, as quantified in **SI Section 4**. To achieve linear dispersion in a nonlocal metasurface with a large $Q$ factor, we dispersion-engineer the q-BIC along the $y$-plane[36,38], while retaining parabolic dispersion along the $x$-plane. The four-ellipse supercell (**Fig. 1b**) collectively supports two orthogonally polarized, zone-folded q-BICs whose folded light cones are centered at opposite reciprocal-lattice vectors. The cones are degenerate in the $k_x$–$\omega$ plane and form a Dirac-like crossing in the $k_y$–$\omega$ plane, forming a linearly dispersed leaky-wave response that enables long-range spatial coherence. All four ellipses participate in the collective modes shown, and they do not form separate local and nonlocal pairs. Their rotations tune the local dipole orientation and geometric phase while weakly perturbing the collective response. The lattice arrangement is detailed in **SI Section 5**.

Previous leaky-wave metasurfaces were generally driven by externally excited coherent guided modes with spin-locked orthogonal polarizations, whose propagation directions were explicitly incorporated into the design of the amplitude, phase and polarization of the outcoupled field[36,38]. In our thermal leaky-wave metasurface, the leaky-wave modes are populated by random fluctuations, which excite the two spin-locked counterpropagating modes equally: the $+k_y$ mode emits left-handed circularly polarized (LCP) light, while the $-k_y$ mode emits right-handed circularly polarized (RCP) light. The radiation from each branch can be modeled using the corresponding coherently excited eigenmode, while the two thermally populated branches are combined incoherently with equal weights because of their mutual incoherence (**SI Section 6**). Normal to the metasurface, the two modes are degenerate, resulting in unpolarized emission. Our experimental results show a quality factor of ~180 and an emissivity of ~0.6 (**Fig. 1c**), and angular emission within ~0.67° and 11° along the *y*- and *x*-axes, respectively (**Fig. 1d**), corresponding to coherence lengths[6] of ~$85\lambda_0$ and $5.2\lambda_0$, respectively. These values show that dispersion engineering enables dramatically enhanced coherence lengths. We benchmark our thermal leaky-wave metasurface—which simultaneously achieves long-range coherence and high emissivity—against previously reported thermal emitters (**SI Table S1**) [6,7,12,17,19,20,23,42,45–51]. The resulting performance is comparable to that of prior strictly periodic structures, yet those offer only limited control over the thermally emitted wavefront. Here, we achieve these same features with fully aperiodic metasurfaces, designed to impart a spatially varying vectorial phase profile onto the emitted light (**Fig. 1a**).

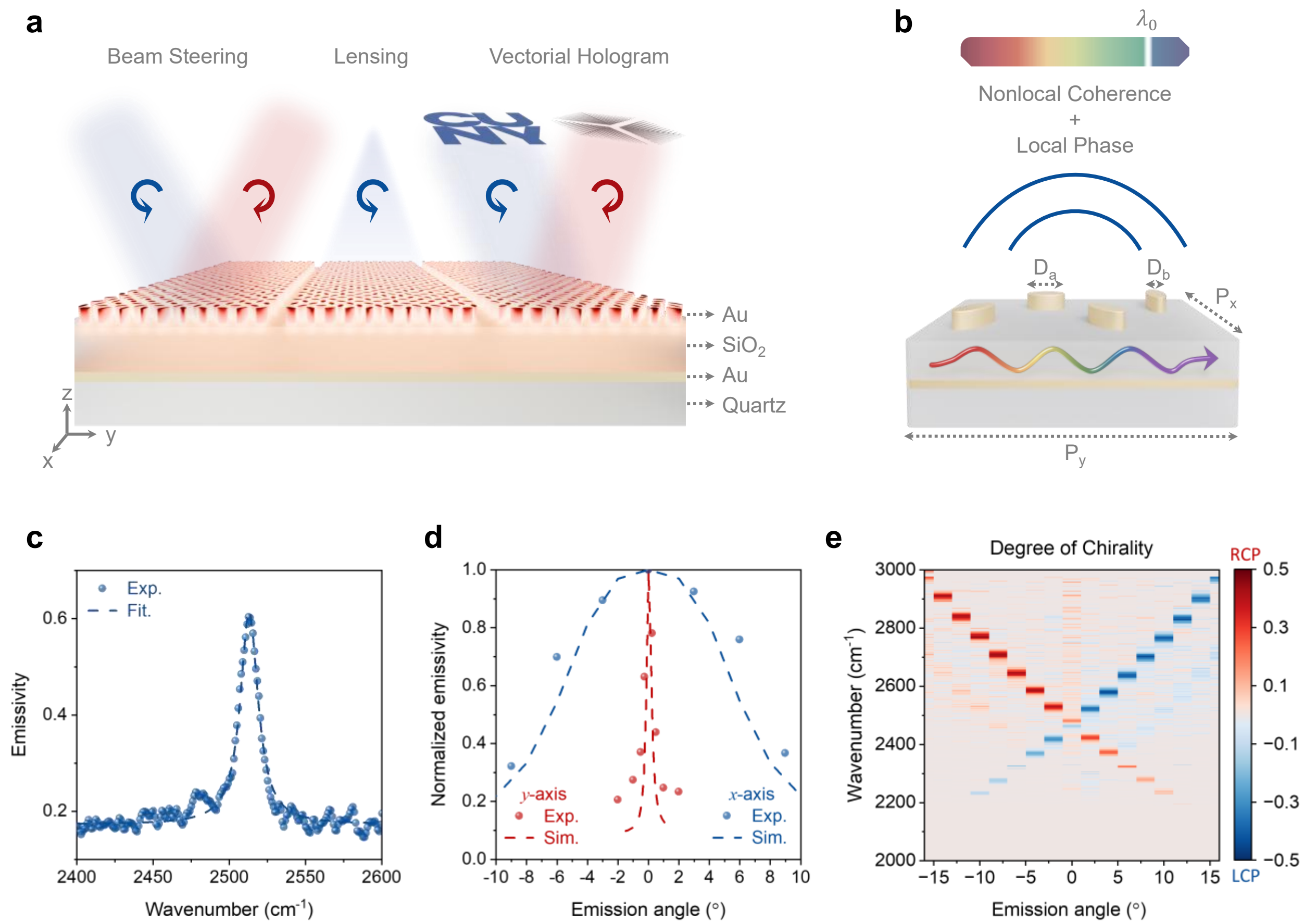

**Fig. 1. Thermal leaky-wave metasurface producing spatially and temporally coherent thermal emission with wavefront control. a,** Conceptual view of our thermal metasurface. The thermal emission can be shaped into programmed wavefronts, including beam steering, focusing, and vectorial holograms. **b,** Schematic of the leaky-wave metasurface unit cell. The unit cell can be individually adjusted to control the local phase, while a collective resonance across the lattice provides coherence. $D_a$ (700 nm) and $D_b$ (400 nm) are the ellipse diameters along the long and short axes, respectively. $P_x$ (1650 nm) and $P_y$ (3300 nm) are the periodicities along the two in-plane axes. The thickness of the $SiO_2$ spacer is 1860 nm. **c,** Measured emissivity spectrum of a periodic leaky-wave metasurface, featuring large temporal coherence. **d,** Angular emission along the *x*- and *y*-axes. The data points were obtained experimentally, and the dashed curves were obtained from finite-element simulations. **e,** Spin-momentum-locked thermal emission manifested by chiral selectivity in the *yz* plane. The degree of chirality is defined as $(\varepsilon_{RCP}-\varepsilon_{LCP})/(\varepsilon_{RCP}+\varepsilon_{LCP})$. The plot is post-processed to remove the influence of background emission, and the raw data are shown in **SI Section 1**. Notably, the experimental emissivity (peak value of 0.6) suffers from convolution effects (**SI Section 17**). We estimate a peak emissivity of ~0.7 from the convolution calculation.

## Beam Steering of Thermal Emission with Preserved Coherence

While the demonstration in **Fig. 1** features a periodic metasurface, the proposed platform is ideally suited to demonstrate wavefront engineering through local geometric phase control. In our leaky-wave metasurface design, the four metaunits within each unit cell are rotated by the same angle (**Fig. 2a**), tailoring the phase of the emitted light unit cell by unit cell. The proper choice of preserved symmetry within each unit cell allows periodicity to be broken without impacting the overall coherence of thermal emission. This technique has been used in past demonstrations to realize metasurface lenses and holograms driven by external coherent light sources, i.e., lasers and radio frequency sources[36,38]. However, the incoherent nature of thermal emission requires more sophisticated control over the preserved nonlocal nature of the metasurface. We first examine the local geometric phase imparted within each unit cell as we vary the angle $\theta_{\mathrm{BIC}}$ in **Fig. 2a**. The local geometric phase $\Phi_{\mathrm{geo}}$ varies linearly with $\theta_{\mathrm{BIC}}$ and spans a complete $2\pi$ phase range, while the emissivity remains around 0.9, as shown in **Fig. 2b**. Importantly, the $\Phi_{\mathrm{geo}}$ values for the RCP and LCP modes are opposite, due to their conjugate relationship, and in our metasurface we expect both RCP and LCP modes to be equally excited because of symmetry.

With the ability to locally control $\Phi_{\mathrm{geo}}$, our first demonstration focuses on a phase-gradient metasurface, aiming at steering directional thermal emission to a desired angle (**Fig. 2c**). The angle $\theta_{\mathrm{BIC}}$ is varied linearly along the *y*-axis, as shown in the representative SEM image (**Fig. 2a**), forming a constant phase gradient that controls the emission angle. We constructed a linear phase gradient leaky-wave metasurface with a phase period spanning six unit cells, corresponding to thermal emission expected to be centered at 11.3˚. Indeed, our experimental results show thermal emission at ~2500 $cm^{-1}$ steered at 12° from the normal for both circular polarizations (**Fig. 2d, e**), consistent with our numerical simulation (**SI Section 7**). Because RCP and LCP modes experience opposite $\Phi_{\mathrm{geo}}$, LCP and RCP emissions are steered symmetrically toward opposite angles (~12˚ and ~-12˚). Critically, the spatial and temporal coherence of the thermal emission are maintained despite the broken periodicity (**SI Section 8**), indicating that the local perturbations do not impact the mode nonlocality, and the band structure remains intact, even though the mode leakage is locally customized. This framework therefore differs from our previous work[23], where the coherence properties were not confirmed after breaking periodicity, and holds promise for advanced functionalities such as lensing and holography.

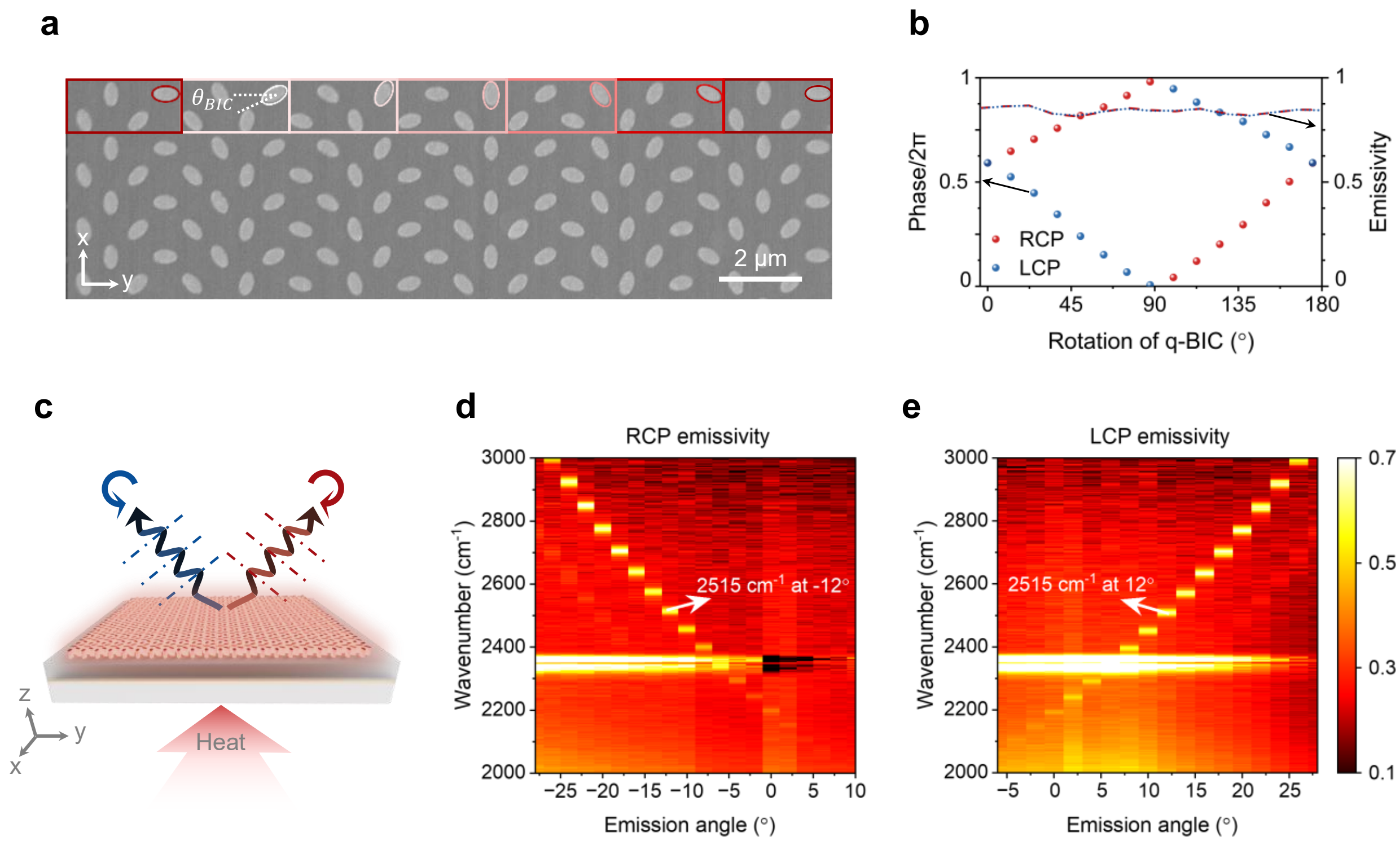

**Fig. 2. Beam steering of thermal emission by local phase manipulation. a,** A representative SEM image of the realized metasurface. The metaunits in a unit cell, outlined by the solid rectangle, are continuously rotated through six periods along the *y*-axis to generate a full 0–2$\pi$ phase gradient. This progression is visualized by a bright-to-dark color code. **b,** $\Phi_{geo}$ for LCP and RCP excitations versus metaunit rotation angle ($\theta_{BIC}$). The emissivity is maintained while the phase varies. **c,** Conceptual view of thermal beam steering. Two chiral states, emitting close to the blackbody limit, are steered to customizable opposite angles. **d, e,** Measured RCP and LCP thermal emission. The strong feature at ~2350 cm$^{-1}$ stems from $CO_2$ absorption in the atmosphere.

## Cylindrical Thermal Focusing

After demonstrating thermal beam steering, we extend our demonstration to an aperiodic leaky-wave metasurface: a thermal metalens using a rigorous hyperbolic phase profile to focus its emission (**Fig. 3a(i)**). We designed a cylindrical lens featuring a line focus along the *y*-axis, with linear dispersion and a long coherence length. For LCP emission, the cylindrical phase profile is $\Phi_y = -k_0\sqrt{(f^2 + y^2) - f}$, with a focal length of 50 mm and an aperture of 5 mm. The corresponding response is modeled analytically using spatiotemporal coupled-mode theory (STCMT)[33] (**Fig. 3b**), which allows us to define a modal band structure for such aperiodic field profiles. Experimentally, we measured the emission by moving the sample away from the collection plane, confirming focused thermal emission approximately 50 mm from the sample, as designed, and in good agreement with our analytical model (**Fig. 3c**). As a result of the linear dispersion imprinted into the underlying nonlocal modes, the spectrally resolved focal-plane response shifts laterally along the *y*-axis within the measured bandwidth, as confirmed by both STCMT and experiment (**Fig. 3d, e**). This feature can potentially be used for spectrometry applications.

Notably, the thermal emission from our metasurface is spin–momentum-locked, and only LCP emission is focused. By contrast, the opposite phase imparted to the RCP emission produces a diverging wavefront, as illustrated schematically in **Fig. 3a(ii)**. This corresponds to a virtual focus 50 mm behind the sample, as predicted by STCMT (**Fig. 3f**) and confirmed experimentally (**Fig. 3g**). Spin-momentum locking preserves the degree of chirality, even though both channels radiate structured waves: at the LCP focal spot, the RCP channel diverges, and the experimentally measured and analytically calculated degrees of chirality are ~-0.5 and -0.75, respectively. The degradation of the experimentally measured degree of chirality is caused by the unpolarized and nonresonant background emission (**SI Section 10**). As a result of spin-momentum locking and linear dispersion, the intensities of the two opposite spins at various wavelengths are mapped to distinct, linearly displaced positions in the front and back focal planes, suggesting applications for chiral-resolved spectrometry, as schematically shown in **Fig. 3a**.

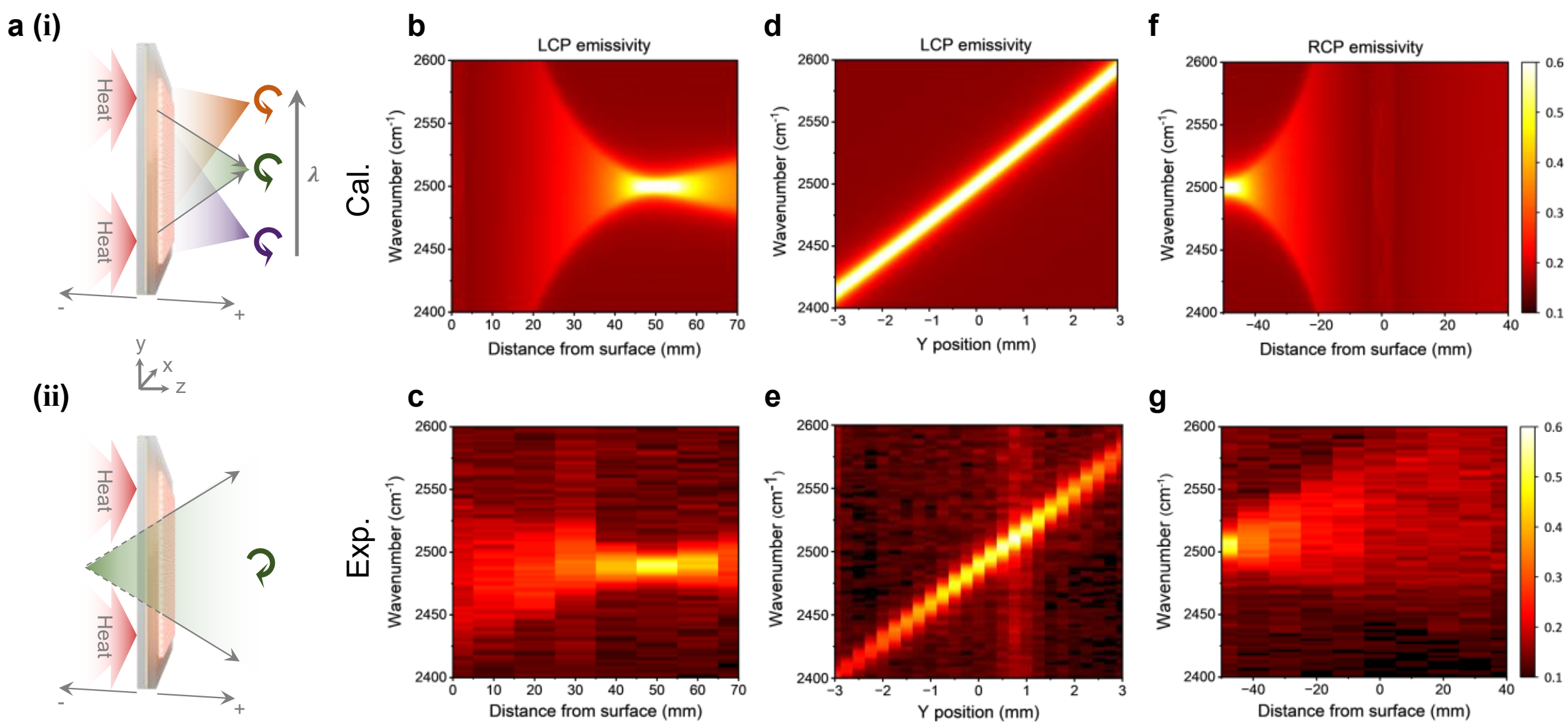

**Fig. 3. Spin-momentum-locked cylindrical thermal lens. a,** Schematic of the measurement and concept. **(i) LCP** emission is focused in front of the surface (positive distance), with a spectrally resolved lateral displacement at the nominal focal plane. **(ii)** RCP emission diverges, with a virtual focus behind the surface (negative distance in panels **f, g**). **b, c,** Focusing in real space for LCP emission, with a focal distance of 50 mm from the metasurface. **d, e,** Linear dispersion of LCP emission along the *y*-axis in the focal plane. The LCP focal spot has an FWHM of ~750 µm, and the spectrally resolved response shifts laterally at the nominal focal plane. Since the sample size (5 mm) is significantly larger than the coherence length (~340 µm), the collected emission is partially coherent, preventing a diffraction-limited spot size, as discussed in detail in **SI Section 9**. For a leaky-wave metasurface comparable in size to or smaller than the coherence length, the spot size would approach the diffraction limit (**Supplementary Fig. S11**). **f, g,** Divergent RCP emission, with a virtual focal spot located 50 mm behind the surface. Panels (**b, d, f**) show STCMT calculations, whereas panels (**c, e, g**) show the corresponding experimental data.

## Two-dimensional Thermal Focusing

To test simultaneous phase programming along both principal axes, we designed a point-focusing metalens with a two-dimensional hyperbolic phase profile (**Fig. 4a**):

$$\Phi(x, y) = -k_0 \left[\sqrt{f^2 + x^2 + y^2} - f\right]$$

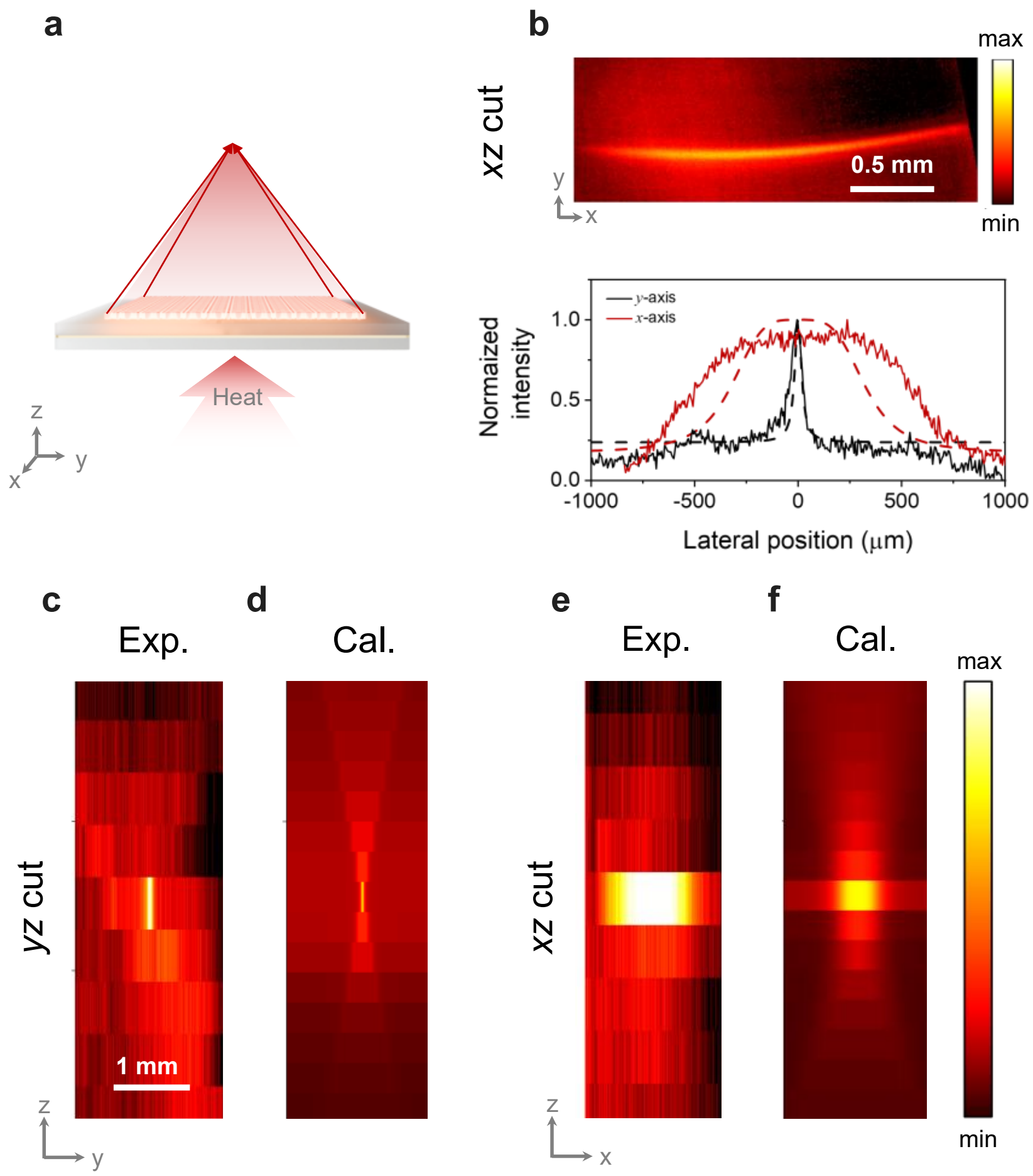

**Fig. 4. Two-dimensional thermal metalens measured in the focal and axial planes and modeled with two-dimensional STCMT. a,** Device concept. The lateral sizes of the lens are 2.6 mm and the designed focal length is 4 mm. **b,** Experimental focal-plane intensity map and *x*- and *y*-direction line profiles through the focal maximum; solid curves show the experimental data, and dashed curves show the corresponding STCMT comparison. The measured (calculated) FWHM along the two principal axes are 1115 (620) and 67 (40) µm, respectively. **c, d,** Experimental and STCMT *yz* axial sections, respectively. **e, f,** Experimental and STCMT *xz* axial sections, respectively. The bottom of the image is ~1 mm from the sample surface. All data here are extracted or calculated with one circular polarization at 2390 cm$^{-1}$, and they share the same lateral scale bar.

At the designed focal plane, the metalens produces simultaneous confinement along both *x* and *y*, demonstrating genuine two-dimensional thermal focusing (**Fig. 4b**). Due to different dispersions and coherence lengths along the two principal axes, the measured focal-plane intensity is not circular; instead, it exhibits a crescent-like profile. Line cuts along the two principal axes confirm focusing in both directions, with a broader focal width along *x* (1115 µm) and a narrower width along $y$ (67 μm). The corresponding FWHM values differ by approximately a factor of 16, consistent with the ratio between the spatial coherence lengths of $5.2\lambda_0$ along $x$ and $85\lambda_0$ along $y$.

Longitudinal intensity maps in the $xz$ and $yz$ planes further confirm convergence toward the designed focal distance along both axes and agree well with the corresponding STCMT predictions (**Fig. 4c–f**). Finally, spectrally resolved measurements show that the focal profile evolves with frequency because the anisotropic dispersion maps different frequencies to different lateral positions along $x$ and $y$, as detailed in **SI Section 11**.

### Vectorial Thermal Hologram

Finally, we demonstrate programmable control of the spatial structure of thermal emission by imprinting a spatially varying vectorial phase profile in our thermal metasurface. Since our metasurface emits spin-momentum-locked thermal light, we exploit this property to realize a thermal vectorial hologram, i.e., an image featuring a spatially continuous distribution of thermal polarization states, as shown in **Fig. 5a**. Specifically, we targeted the LCP and RCP emission channels to form the opposite halves of the silhouette of the Empire State Building at a distance of 16 mm from the surface (**Fig. 5a**). The commonly employed Gerchberg–Saxton algorithm[52] can only optimize one channel, i.e., half of the vectorial hologram, leaving the other chirality unconstrained. We therefore used a stochastic-gradient-descent algorithm[53] to co-optimize the RCP and LCP projections under their shared phase constraint. Despite inherent limitations of spin-momentum-locked phase profiles, the simulated profiles reproduce helicity-dependent image features (**Fig. 5c**), wherein a thermal vectorial hologram emerges at the desired plane with spatial variation in the local degree of chirality. As expected, at different distances from the surface (**Fig.**

**5b, d**), the hologram is distorted, similar to prior demonstrations based on coherent excitations[54]. The inverse-design workflow and numerical calculations are described in **SI Section 12**, and the algorithm is openly available[55].

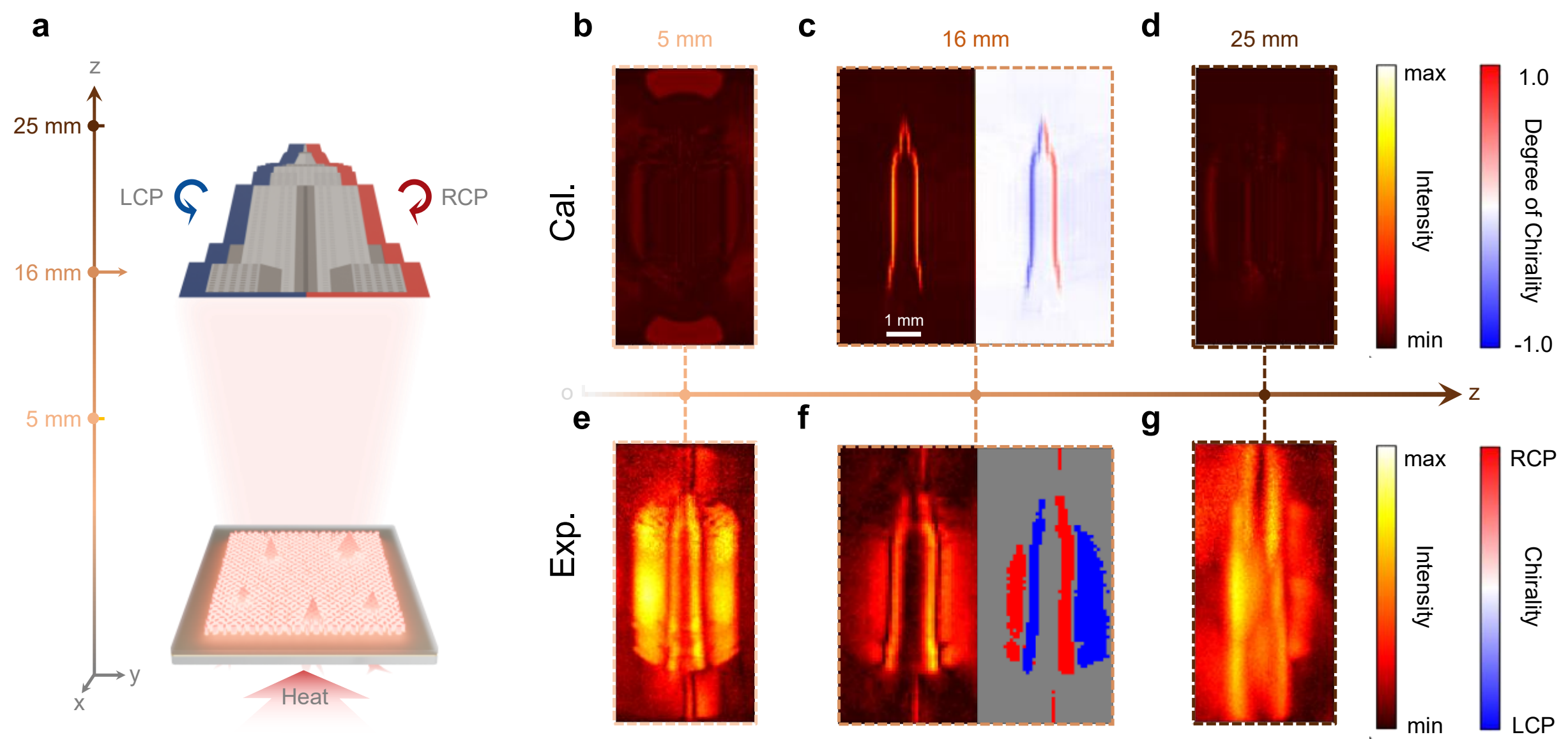

**Fig. 5. Polarization-resolved thermal hologram. a,** Device and polarization schematic. The lateral sizes of the sample are 8 mm and the designed working distance is 16 mm. **b–d,** Calculated intensity and chirality-dependent response at selected propagation distances. **e–g**, Corresponding measured intensity, chirality, and polarization-resolved maps. In the experimental data, the direct TWINS reconstruction provides signed third Stokes parameter, *S3*, and the handedness is assigned from the complex Fourier phase and cross-validated against the separately acquired polarization-resolved RCP and LCP maps (**SI Section 13 and Section 14**). Experimental camera counts and calculated emissivity are compared through spatial form and normalized profiles rather than absolute magnitude. All data here are extracted or calculated at 2544 $cm^{-1}$, and they share the same lateral scale bar.

Based on the designed phase profile, we fabricated the thermal holographic metasurface and measured its performance. The experimental data reproduce the main spatial features of the calculations (**Fig. 5e–g**), including opposite handedness in the two arms of the Empire State Building and the expected evolution with propagation distance. As the emission is linearly dispersive, changing the wavelength laterally shifts the reconstructed image, as shown in **SI Section 15**. Although the optimized phase mask is aperiodic (**Fig. S18**) and the spatial coherence differs along $x$ and $y$ axes, the metasurface still supports a collective resonance across its aperture, forming a carefully designed image.

This demonstration provides programmable spatial control over thermal light, while retaining the measured temporal and spatial coherence and high emissivity. Although the two conjugate fields (LCP and RCP) are emitted simultaneously with opposite phase profiles, our co-optimization strategy shows that this fundamental limitation, stemming from the single-layer nature of our metasurface, can be effectively mitigated to realize a fully vectorial hologram. This is achieved using a common, CMOS-compatible material platform and standard nanofabrication techniques, offering a compact and integrable platform for custom thermal wavefront generation.

## Conclusion

In this work, we report a thermal metasurface demonstrating programmable wavefront shaping of spin–momentum-locked thermal emission. With a $Q$ factor of ~180, coherence lengths of $85\lambda_0$ and $5.2\lambda_0$ along the two in-plane axes, and an experimental emissivity larger than 0.6, our metasurface exhibits resonant properties comparable to those previously reported for periodic thermal-emission structures while additionally enabling aperiodic programming of the emitted wavefront in amplitude, phase, and polarization. We demonstrated beam steering, cylindrical and two-dimensional focusing, and vectorial holography. In particular, the experimentally demonstrated thermal vectorial hologram highlights the feasibility of spatially varying wavefront shaping for thermal emission: even under spin–momentum locking, the two spin-locked channels can be co-optimized under their shared phase constraint. Together, these results establish a leaky-wave platform for programmable thermal wavefront shaping. Future work may test whether these dispersion and phase-control concepts can be adapted to luminescent emitters[56,57] and fluorescence[58], as well as to coherent processes such as lasing, nonlinear optics, and quantum emission.

## Methods

### Fabrication

We fabricate our metasurfaces using standard lithography and metal lift-off processes. The substrate preparation is as follows: First, a 100 nm layer of gold, with a 10 nm chromium adhesion layer, is deposited onto a quartz substrate via e-beam evaporation. This is followed by the deposition of a silicon dioxide layer (1860 nm) using plasma-enhanced chemical vapor deposition (PECVD). The substrate is then coated with polymethyl methacrylate (PMMA) resist, and the metasurface pattern is defined using e-beam lithography (Elionix 50 keV). To mitigate charging issues during lithography, an anti-charging layer (DisChem DisCharge H2O) is spun onto the PMMA. After lithography, the DisCharge is removed by rinsing with DI water, and the resist is developed using methyl isobutyl ketone (MIBK). Finally, chromium (5 nm) and gold (50 nm) are deposited via e-beam evaporation to form the metasurface layer. The process concludes with Remover PG immersion, lifting off the unexposed areas and transferring the pattern from the PMMA to the gold layer.

### Measurements

Measurements for **Fig. 1–3** used an external three-axis translation and rotation stage; the FTIR beam paths are shown in **SI Section 16**. Systematic and random uncertainties are discussed in **SI Section 17**. For the cylindrical lens measurements, we fixed the collection optics and raster-scanned the sample position to acquire polarization-resolved hyperspectral data as a function of $x$, $y$, $z$, and wavenumber. The circular lens and vectorial hologram in **Figs. 4** and **5** were measured with a Translating-Wedge-based Identical Pulses eNcoding System (TWINS) hyperspectral imager, described in **SI Section 13**.

The free-space optical paths in the measurement setups described above were neither enclosed nor purged with nitrogen, and spectral features near the atmospheric $CO_2$ absorption band are therefore contaminated. Enclosing and purging the complete path would suppress these artifacts.

For FTIR measurements, sample emission was collected by the external optics and detected with a mercury-cadmium-telluride (MCT) detector. To calculate the emissivity, thermal emission measurements were collected from our device at 350˚C. These measurements were then compared to the thermal emission measured from an emissivity standard at the same temperature and angle of emission. We used 500 µm tall vertically aligned carbon nanotubes (VACNTs) grown on a Si substrate, grown by Nanotechlabs Inc., as an emissivity standard ($\epsilon$~0.97). These VACNTs provide a high, consistent value for emissivity that is spectrally flat throughout the *IR* and stable with temperature. The signal collected by the MCT detector in these measurements contains the emission from both the sample as well as the internal optics of the FTIR.

$$M(T_{sample}, T_{ambient}, \lambda, \theta) = R(T_{ambient}, \lambda)\left[S(T_{sample}, \lambda, \theta) + G(T_{ambient}, \lambda)\right] \qquad \text{Eq. (1)}$$

Here $M$ is the total measured signal, $R$ is a response function for the internal and external optics, $S$ is the signal originating from the sample and $G$ is the 'background' emission from the internal optics. Thus, in order to isolate the signal from the sample, a background measurement was taken by replacing the sample with a gold mirror. The resultant spectrum is a product of the response function *R* with the background emission *G*. Once the sample, emissivity standard and background emission have been measured we can rearrange equation (1).

$$\frac{M(T_{sample},T_{ambient},\lambda,\theta)-R(T_{ambient},\lambda)G}{M(T_{VACNT},T_{ambient},\lambda,\theta)-R(T_{ambient},\lambda)G}=\frac{R(T_{ambient},\lambda)S_{sample}(T_{sample},\lambda,\theta)}{R(T_{ambient},\lambda)S_{standard}(T_{sample},\lambda,\theta)}=\varepsilon(T_{sample},\lambda,\theta)$$

Eq. (2)

**Numerical simulations**

All full-wave simulations except those in **SI Section 6** used the finite-element method (CST Studio 2019). Unit cell boundary conditions were used in all simulations, and the circularly or linearly polarized excitations were selected depending on the specific figures. The numerical simulations in **SI, section 6** are performed with Lumerical FDTD with Bloch boundary conditions with waveguide mode as input, and details are given in the corresponding section.

## Acknowledgements

**Funding**. This work was supported by the Vannevar Bush Faculty Fellowship Program and the Simons Foundation. MG acknowledges the European Union's Horizon Europe research and innovation program under the Marie Skłodowska-Curie UTOPIA HORIZON-MSCA-2023-PF-GF grant agreement No. 101151191. **Author contributions:** M. H. and L. J. contributed equally. M. H., A. O. and A. A. conceived the idea. M. H. designed the devices, performed numerical calculations, characterized the samples with FTIR and analyzed the results. L. J. fabricated the sample and prepared the graphical visualizations. M. G., M. R., and M. H. performed the TWINS hyperspectral imaging. S. Y., S. M. and A. O. performed the STCMT modeling. Y. G. performed numerical calculations. S. G. assisted with the fabrication. A. A. supervised the research. All authors wrote the manuscript. **Competing interests**: The authors declare no competing financial interests. **Data and materials availability**: All data needed to evaluate the conclusions in the paper are available in the main text and the Supplementary Information.